\documentclass[article,a4paper,preprint,sort&compress]{revtex4}

\usepackage{color}
\usepackage{graphicx}
\usepackage{amsmath}
\usepackage{amstext}
\usepackage{units}
\usepackage{epstopdf}

\begin{document}

\title{Characterizing the variation of propagation constants in multicore fibre}
\author{Peter J. Mosley$^{1}$}\email{p.mosley@bath.ac.uk}
\author{Itandehui Gris-Sanchez$^1$, James M. Stone$^1$, Robert J. A. Francis-Jones$^1$, Douglas J. Ashton$^2$, Tim A. Birks$^1$}
\affiliation{$^1$Centre for Photonics and Photonic Materials, Department of Physics, University of Bath, Bath, BA2 7AY, UK \\$^2$Department of Physics, University of Bath, Bath, BA2 7AY, UK}

\begin{abstract}
We demonstrate a numerical technique that can evaluate the core-to-core variations in propagation constant in multicore fibre. Using a Markov Chain Monte Carlo process, we replicate the interference patterns of light that has coupled between the cores during propagation. We describe the algorithm and verify its operation by successfully reconstructing target propagation constants in a fictional fibre. Then we carry out a reconstruction of the propagation constants in a real fibre containing 37 single-mode cores. We find that the range of fractional propagation constant variation across the cores is approximately $\pm2 \times 10^{-5}$.
\end{abstract}

\maketitle

%\ocis{060.2270 Fiber characterization, 060.2280 Fiber design and fabrication, 060.4005 Microstructured fibers}

Multicore fibre (MCF) is finding applications in a number of different areas of science and technology. Spatial-division multiplexing (SDM) with MCF provides a possible route to bypass the impending capacity crunch in optical telecommunication networks \cite{Richardson2013Space-division-multiplexing-in-optical}, with single-fibre capacity now exceeding 1 Pb/s using multicore architecture \cite{Takara20121.01-Pb/s-12-SDM/222-WDM/456, Qian20121.05Pb/s-Transmission-with}. MCF allows a range of new capabilities in nonlinear endoscopy through propagating pulses in separate cores to enable the coherent synthesis of a high-intensity pulse at the sample \cite{Thompson2011Adaptive-phase-compensation, Mansuryan2012Spatially-dispersive-scheme, Andresen2013Two-photon-lensless-endoscope}, or by excitation of functionalised distal core tips. MCF has the potential to revolutionise astronomical instrumentation in the emerging field of astrophotonics \cite{Bland-Hawthorn2009Astrophotonics:-a-new-era-for-astronomical-instruments}. It can be used to reformat light from a telescope's focal plane and guide it to a sensor, to filter out unwanted emission lines originating in the Earth's atmosphere without the need for large monochromators \cite{Bland-Hawthorn2011A-complex-multi-notch-astronomical}, and to interface multimode inputs with devices that require single-mode operation \cite{Birks2012Photonic-lantern-spectral}. In quantum optics, coupled-core MCFs provide a platform for extending the dimensionality of quantum walks of single and entangled photons beyond the capabilities of planar \cite{Peruzzo2010Quantum-Walks-of-Correlated} or direct-write waveguide technology \cite{Owens2011Two-photon-quantum-walks, Poulios2014Quantum-Walks-of-Correlated}. MCF can also be used in optical switching and modelocking applications \cite{Nazemosadat2014Design-considerations-for-multicore}.

However, the application of MCF will always be limited by the precision that can be achieved in its fabrication. For example, if MCF is to be used as a filter for narrowband spectral features, Bragg gratings must be written into the cores; non-uniformity in the propagation constant between cores results in a spread in filter wavelength and a corresponding decrease in extinction across the device \cite{Birks2012Photonic-lantern-spectral, Lindley2014Core-to-core-uniformity-improvement}. To carry out a photonic quantum walk in MCF the cores must be strongly coupled; any variation in propagation constant from core-to-core will accumulate phase mismatch across the MCF as the probability amplitude propagates. If this phase mismatch is too great, quantum correlations will no longer be observed at the output of the fibre. Although in the case of SDM data transmission, coupling between the cores is usually undesirable \cite{Koshiba2009Heterogeneous-multi-core-fibers:, Imamura2011Investigation-on-multi-core-fibers}, accurate characterization of the structural uniformity of the fibre would yield important information about its transmission characteristics.

Obtaining information about the inevitable variations in fibre structure that arise during fabrication is not trivial \cite{Fini2012Crosstalk-in-multicore-fibers}. Imaging the structure does not yield the required precision or accuracy. Over long fibre lengths where the cores are weakly coupled, information about the coupling strength may be inferred from optical time-domain reflectometry \cite{Yoshida2013Detailed-comparison-between} though this provides no information about variations in propagation constant, and in some applications a sufficiently long length of MCF may not be available.

In this manuscript we present a method of determining the differences in propagation constant in MCF. We achieve this by using a short length of MCF in which the cores are strongly coupled. We measure output intensity distributions for various input fields, and using a Monte Carlo technique to reconstruct numerically the differences in propagation constant between the cores.

\section{Propagation in multicore fibre}

The linear propagation of light at frequency $\omega$ in the $z$-direction in a MCF can be described by the coupled amplitude equation:
\begin{equation}
\frac{d \pmb{A}}{d z} = i \pmb{M} \pmb{A}
\label{eq:propagation}
\end{equation}
where $\pmb{A} = \{A_1 \hspace{2mm} A_2 \hspace{2mm} A_3 \ldots \hspace{2mm} A_m \}$ is the complex amplitude of light in each of the $m$ cores. The transfer matrix $\pmb{M}$ can be written
\begin{equation}
\pmb{M} = \left( \begin{array}{ccccc}
\beta_1	& g_{12}	& g_{13}	&		& g_{1m}	\\
g_{21}	& \beta_2	& g_{23}	& \cdots	& g_{2m}	\\
g_{31}	& g_{32}	& \beta_3	&		& g_{3m}	\\
        		& \vdots  	&		&  \ddots	& \vdots	\\
g_{m1}	& g_{m2}	& g_{m3}	& \cdots	& \beta_m	\end{array}\right),
\end{equation}
where $\beta_i$ is the propagation constant in the $i^\text{th}$ core and $g_{ij} = g_{ji}$ is the coupling strength between the $i^\text{th}$ and $j^\text{th}$ cores. In general the propagation constants and couplings are strong functions of frequency, however we will be considering only narrow wavelength ranges at any one time and therefore the frequency dependence of $\pmb{M}$ has been omitted for notational convenience. The propagation constant in the $i^\text{th}$ core is defined as $\beta_i = n^\text{eff}_i(\omega) \omega/c$, where $n^\text{eff}_i(\omega)$ is the effective refractive index in the $i^\text{th}$ core. The coupling strength $g_{ij}$ is defined as the overlap of the unperturbed mode profiles of the $i^\text{th}$ and $j^\text{th}$ cores, $\bar{\Psi}_i(\omega,x,y)$ and $\bar{\Psi}_j(\omega,x,y)$ respectively, with the perturbation introduced by the refractive index contrast of the $j^\text{th}$ core:
\begin{equation}
g_{ij} = \frac{2\pi}{\lambda_0} \iint \, dx \, dy \, [n(x,y) - \bar{n}_i] \, \bar{\Psi}_i(\omega,x,y) \, \bar{\Psi}_j(\omega,x,y).
\label{eq:general_coupling}
\end{equation}
Here the unperturbed mode profiles obey the normalization condition $ \iint \, dx \, dy \,\bar{\Psi}^2_i(\omega,x,y) = 1$. We model MCF as a nominally regular array of circular cores with step-index refractive index profiles, in which the mode profiles are Bessel functions with exponentially-decaying wings. Due to the exponential decay of the mode profiles, the dynamics of these structures are dominated by nearest-neighbour couplings. Non-nearest-neighbour couplings can be ignored to an excellent approximation and we will adopt this convention henceforth.

For a given wavelength, $\pmb{M}$ describes everything that we require to describe light propagation through the MCF. Given sufficiently accurate knowledge of the MCF structure we could straightforwardly calculate the propagation constants and mode coupling parameters for the fibre. This would allow us to solve the propagation equation \ref{eq:propagation}, either through finding the matrix exponential of $\pmb{M}$ or by a stepwise numerical method, and hence determine the output amplitudes for any number of fibre cores. However, direct measurement of the structure does not provide sufficient accuracy to determine the values of propagation constant and coupling strength.

In a perfect MCF, all the fibre cores would be identical, with radius $a$, centre-to-centre separation $d$ and index contrast with respect to the surrounding cladding material of $\Delta n = n_\text{core} - n_\text{cladding}$. In this ideal situation the propagation constants for each core and the nearest-neighbour coupling strengths would be identical. Light coupling between two cores within the MCF would be perfectly phasematched, and, if light was input to just one core, it would spread symmetrically and eventually re-assemble in phase in the conjugate core after reflecting from the MCF boundaries.

\section{Impact of structural variations}

Any MCF that we might fabricate will not be ideal; the structure will inevitably contain some level of variation. In general this could be either in the longitudinal or transverse direction, corresponding respectively to the $i^\text{th}$ core varying along its length ($\beta_i = \beta_i(z)$) or the propagation constants and coupling strengths varying from core to core ($\beta_i \neq \beta_j$ and $g_{ij} \neq g_{jk}$; note however that for physically realistic systems we always require that the coupling is reciprocal so that $g_{ij} = g_{ji}$ and so on) \cite{Koshiba2011Multi-core-fiber-design}. Variations in propagation constant and coupling strength both have the potential to influence the amplitude of light in each core at the output. Local increases in coupling strength create ``preferred'' routes along which the light spreads, whereas differences in propagation constant  imperfect phasematching between cores resulting in incomplete transfer of light from one core to the next. Although it may seem at first glance that the latter effect is somewhat secondary, in fact -- as we will see -- it typically dominates the output intensity distributions.

We can calculate the relative impact of these effects on the flow of light between the cores by considering the analogous case of a two-mode coupler (equivalent to the simplest MCF containing only two cores). Following the analysis in \cite{Snyder1983Optical-waveguide-theory}, the coupling strength between a pair of identical step-index cores, reduces to
\begin{equation}
g_\text{2-core} = \sqrt{\frac{2 \Delta n}{n_\text{core}}} \, \frac{1}{a} \, \frac{U^2}{V^3} \, \frac{K_0(W d/a)}{K_1^2(W)},
\label{eq:2-core_coupling}
\end{equation}
where $U$, $V$, and $W$ are the usual core, waveguide, and cladding parameters also defined in \cite{Snyder1983Optical-waveguide-theory}, $K_n$ are modified Bessel functions of the second kind, and the other symbols have the meanings previously defined in the text. The beat length between the two cores is defined as $z_b = 2\pi/g_\text{2-core}$. The propagation constants and the parameters $U$, $V$, and $W$ for the unperturbed individual cores are found by solving numerically the eigenvalue equation in the normal manner. Note that the dependence of the coupling strength on both $d$ and $a$ is to a good approximation exponential when the cores are well-separated ($a \ll d$) but the dependence on $a$ becomes less straightforward as the cores are brought into closer proximity due to the dependence on $a$ of many of the parameters in Eq \ref{eq:2-core_coupling}.

\begin{figure}[h]
\centering
\includegraphics[width = 0.7\textwidth]{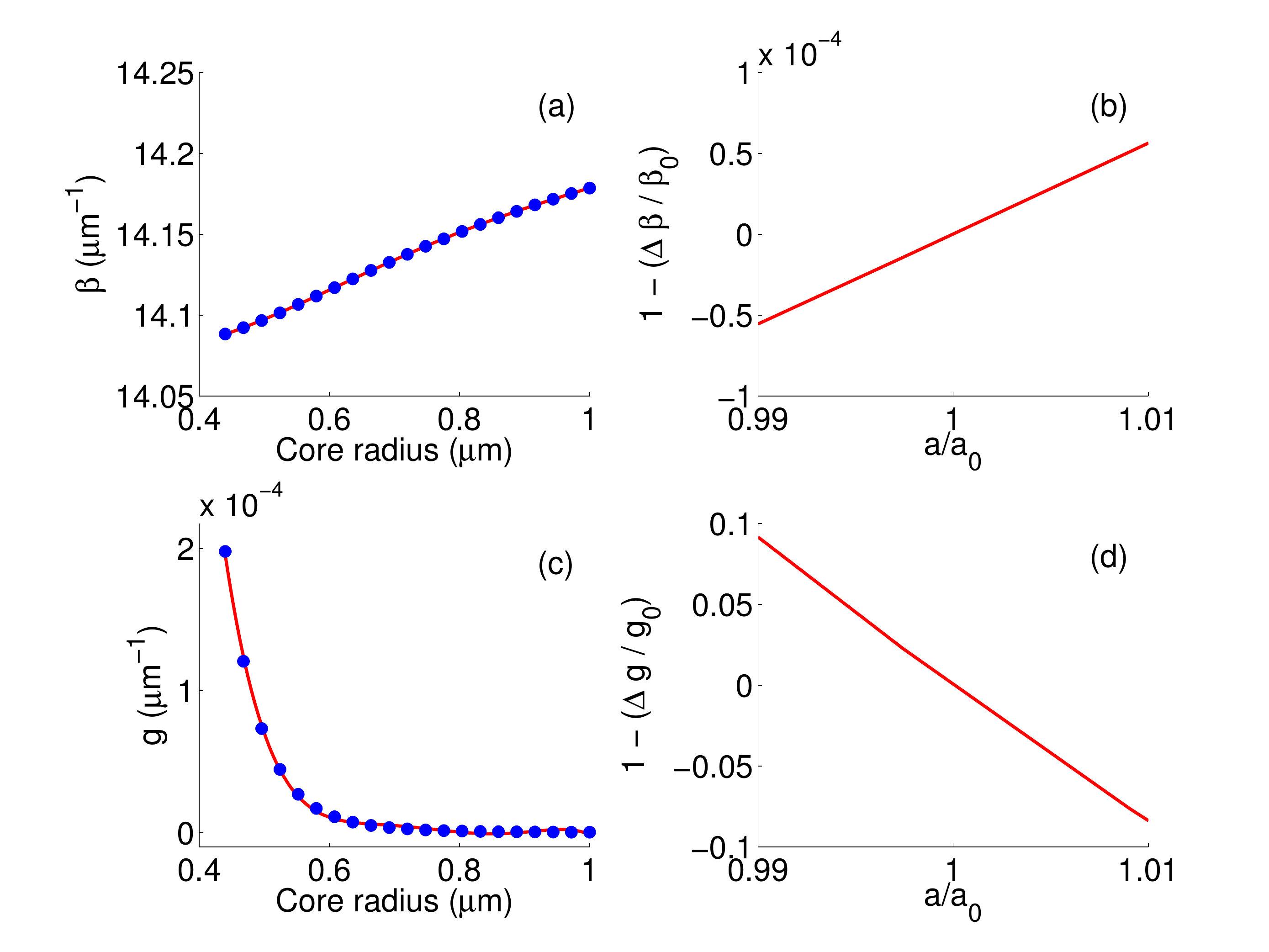}
\includegraphics[width = 0.5\textwidth]{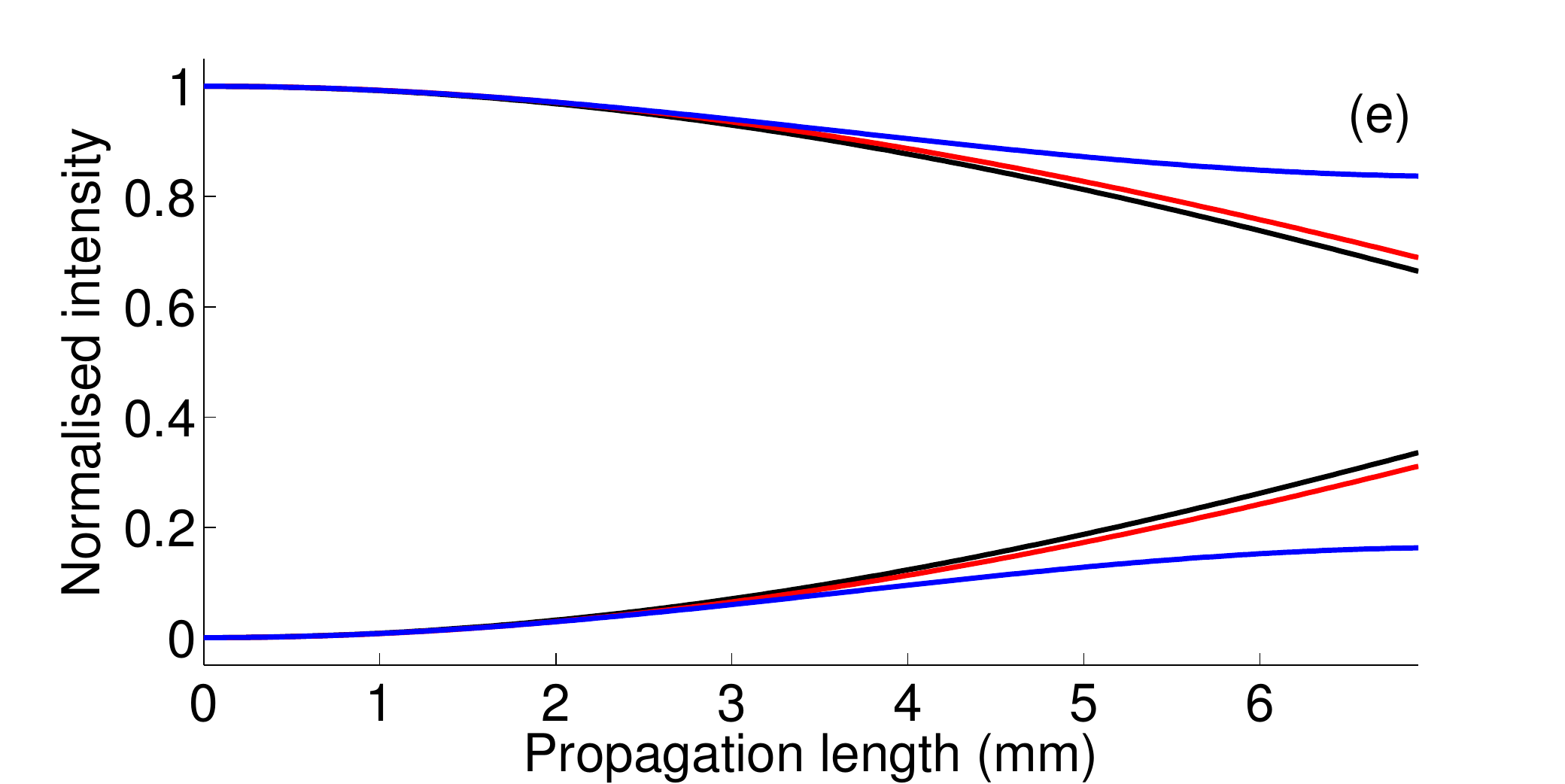}
\caption{Dependence of the propagation constant of an individual core (a, b) and the coupling strength between two identical cores (c, d) on core radius. (b, d) show the fitted dependencies over the typical range of variation in $a$ seen in our MCF. (e) Evolution of intensity for propagation in two coupled cores: identical cores with radii $a_0$, propagation constants $\beta_0$ and coupling $g_0$ in black; identical cores with radii $a_0 +\Delta a$, identical propagation constants $\beta_0 + \Delta \beta$, and modified coupling strength $g_0 - \Delta g$ in red; two different cores with different propagation constants $\beta_0$ and $\beta_0 + \Delta \beta$ but coupling $g_0$ in blue. Parameters were $a_0 = 0.48\,\mu$m, $a_0 + \Delta a = 0.4824\,\mu$m, $d = 8.0\,\mu$m, $\Delta n = 0.02$.}
\label{fig:two_core}
\end{figure}

The relative impact of variations in $\beta$ and $g$ can be found by examining the response of the two-mode coupler to changes core radii. $\beta$ and $g_\text{2-core}$ are plotted for a range of core radii in Figure \ref{fig:two_core}. For parameters similar to those of the fabricated fibre presented later in this work ($a_0 = 0.48\,\mu$m, $d = 8.0\,\mu$m, $\Delta n$ = 0.02), we see that a change in core radius of 1\% yields a fractional change in propagation constant $\Delta \beta/\beta \approx 5\times10^{-5}$ and a fractional change in coupling strength of $\Delta g/g \approx 0.08$. We then solve Equation \ref{eq:propagation} for three situations: two identical cores of nominal radius $a_0$ with coupling $g_0$ and propagation constants $\beta_0$; two identical cores of radius $a_0 + \Delta a$ with identical propagation constants $\beta_0 + \Delta \beta$ but reduced coupling strength $g_0 - \Delta g$; and two slightly different cores of radii $a_0$ and $a_0 + \Delta a$ with different propagation constants $\beta_0$ and $\beta_0 + \Delta \beta$ but coupling $g_0$. For each system we input light to one core only and calculate the fraction that couples into the other after a fixed propagation length. The results of this are shown in Figure \ref{fig:two_core}. It can be seen that in the parameter range of interest, the change in output intensity is dominated by the differences in propagation constant rather than those in the coupling strength.

\section{Reconstruction algorithm}

We would like to find the variations in propagation constant and coupling strength in MCF that result from small structural fluctuations introduced during the fabrication process. To do so we make the following assumptions: each core can be individually addressed at the input end of the MCF; the intensity of light exiting each core of the MCF can be measured; deviations of the output intensity pattern from the ideal are dominated by differences in propagation constant rather than coupling strength (i.e. all nearest-neighbour coupling strengths are assumed equal); non-nearest-neighbour couplings are zero; the MCF is sufficiently short that the fibre length is less than the beat length and the fibre properties can be considered constant in the longitudinal direction (i.e. transverse core-to-core variations dominate); and each core is single-moded over the wavelength range of interest. Hence to describe propagation in a MCF with $m$ cores we need to find a single nearest-neighbour coupling strength $g$ and $m$ values of $\beta_i$.
By addressing each core individually we construct a set of $m$ normalised input intensity patterns $\{\pmb{P}_\text{in}^{(k)}\}$, $1 \leq k \leq m$ with related input amplitudes $\{\pmb{A}_\text{in}^{(k)}\}$, each of which has $m$ elements containing a single nonzero entry equal to unity. These states propagate through a length $L$ of MCF according to Equation \ref{eq:propagation} and form a corresponding set of $m$ output states $\{\pmb{A}_\text{out}^{(k)}\}$, for which we can measure the power exiting each core to yield $m$ output intensity patterns $\{\pmb{P}_\text{out}^{(k)}\}$ with elements $P_{\text{out},i}^{(k)} = | A_{\text{out},i}^{(k)}|^2$. We need to find the $\pmb{M}$ that relates $\{\pmb{P}_\text{in}^{(k)}\}$ to $\{\pmb{P}_\text{out}^{(k)}\}$.

Due to the phase insensitivity of our measurement, the correspondence between any individual pair of input intensity $\pmb{P}_\text{in}^{(l)}$ and output intensity $\pmb{P}_\text{out}^{(l)}$ is insufficient to define $\pmb{M}$ uniquely. Fortunately, the additional information gained from measuring the output intensities related to all $m$ input states provides sufficient constraints to determine all $m$ values of $\beta_i$ contained in $\pmb{M}$, however it results in a set of simultaneous equations that cannot be solved analytically. Furthermore the problem of minimising the difference between the input and output intensities is not generally convex in each individual $\beta_i$; local minima often do not correspond to the global minimum. Hence to reconstruct $\pmb{M}$ we have implemented an iterative Monte Carlo method known as simulated annealing that is typically applied to multi-parameter optimization problems in condensed matter physics \cite{Kirkpatrick1983Optimization-by-simmulated-annealing}.

We know $\{\pmb{P}_\text{in}^{(k)}\}$ and $\{\pmb{P}_\text{out}^{(k)}\}$ for a MCF with unknown transfer matrix $\pmb{M}$. We can approximate the mean nearest-neighbour coupling strength $\tilde{g}$ and mean propagation constant $\beta_0$ of the MCF cores from our imperfect knowledge of the average MCF structure. We construct a transfer matrix $\tilde{\pmb{M}}$ that contains $\tilde{g}$ and a set of randomly-selected $\{\tilde{\beta}_i\}$ that vary around $\beta_0$. The input states are propagated subject to $\tilde{\pmb{M}}$ using Equation \ref{eq:propagation} to yield a set of output intensities $\{\tilde{\pmb{P}}_\text{out}^{(k)}\}$. The difference between the calculated output intensities and the measured output intensities is then found using
\begin{equation}
\mathcal{F} = \frac{1}{2m}\sum_{k=1}^m \sum_{i=1}^m | \tilde{\pmb{P}}_\text{out,i}^{(k)} -  \pmb{P}_\text{out,i}^{(k)} |.
\label{eq:fom}
\end{equation}
This provides a normalised figure of merit that expresses per input state how much power on average ends up in the ``wrong'' core at the output. We then implement a Markov Chain Monte Carlo (MCMC) routine with acceptance criteria inspired by the Metropolis algorithm to reconstruct the differences in the propagation constants \cite{Metropolis1953Equation-of-state-calculations}. We randomly perturb the $\{\tilde{\beta}_i\}$ and repeat the propagation; if $\mathcal{F}$ decreases the new $\{\tilde{\beta}_i\}$ are accepted. However, even if $\mathcal{F}$ increases the move can still be accepted with a small probability to encourage the solution not to become stuck in local minima. The probability of accepting a poor move is exponentially dependent on a ``temperature'' function, $T(n)$, that decreases with the number of iterations $n$, and the difference in the figure of merit between the current position and the new position, $\Delta\mathcal{F}$:
\begin{equation}
P = c_3 n_\text{st} \exp{\left(-\frac{\Delta \mathcal{F}}{T}\right)}, \hspace{5mm} \Delta\mathcal{F} > 0.
\end{equation}
$c_1 \ll 1$ is a constant and $n_\text{st}$ is an integer that increments on every iteration for which the solution is stationary. We choose a temperature function that decreases exponentially with iteration number:
\begin{equation}
T(n) = \frac{1}{2m}\left(c_2 + \exp{\left[-\left(\frac{n}{c_3 N}\right)^2\right]}\right),
\label{eq:temperature}
\end{equation}
where $c_2$ and $c_3$ are constants, and $N$ is the total number of iterations. The MCMC routine is repeated for $N$ iterations during which the magnitude of the random perturbations in $\{\tilde{\beta}_i\}$ is reduced in proportion to the values of $T(n)$ and $\mathcal{F}$. The constants $c_1 - c_3$ are set to obtain optimum performance.

After the algorithm has finished, we are left with a set of reconstructed $\{\beta_i^\text{(rec)}\}$. Note that, although the output intensity patterns give us information about the relative phase of light that has propagated through each core, there is no sensitivity to global phase. Therefore, the reconstruction cannot give any indication of the absolute magnitude of the mean value of $\{\beta_i^\text{(rec)}\}$, only the differences between the individual $\beta_i^\text{(rec)}$. Hence we rescale the $\{\beta_i^\text{(rec)}\}$ to have a mean value equal to $\beta_0$; these then represent our best estimate of the propagation constants in the $m$ cores. We also obtain a figure of merit for the fit, the residual value of $\mathcal{F}^\text{(rec)}$, and the reconstruction can be repeated to find the value of the coupling constant, $g^\text{(rec)}$, that yields the smallest residual value.

\section{Simulated reconstructions}

To test the performance of our algorithm we reconstructed fictional MCFs. We defined a set of target propagation constants $\{\beta_i^\prime\}$ containing both random and systematic variations over a range $\Delta \beta$ and found the associated output intensities $\{\pmb{P}_\text{out}^{\prime (k)}\}$ assuming a uniform coupling strength $g^\prime$ in Equation \ref{eq:propagation}. A random variation of approximately 1\% was added to the individual elements of $\{\pmb{P}_\text{out}^{\prime (k)}\}$ to simulate the effects of measurement noise. We used these output intensities in the algorithm outlined above to reconstruct a set of $\{\beta_i^{\prime \text{(rec)}}\}$ and compared them with the known values of $\{\beta_i^\prime\}$. For each set of propagation constants we ran the reconstruction algorithm 100 times to build up statistics on the quality of the reconstruction.

\begin{figure}[h]
\centering
\includegraphics[width = 0.8\textwidth]{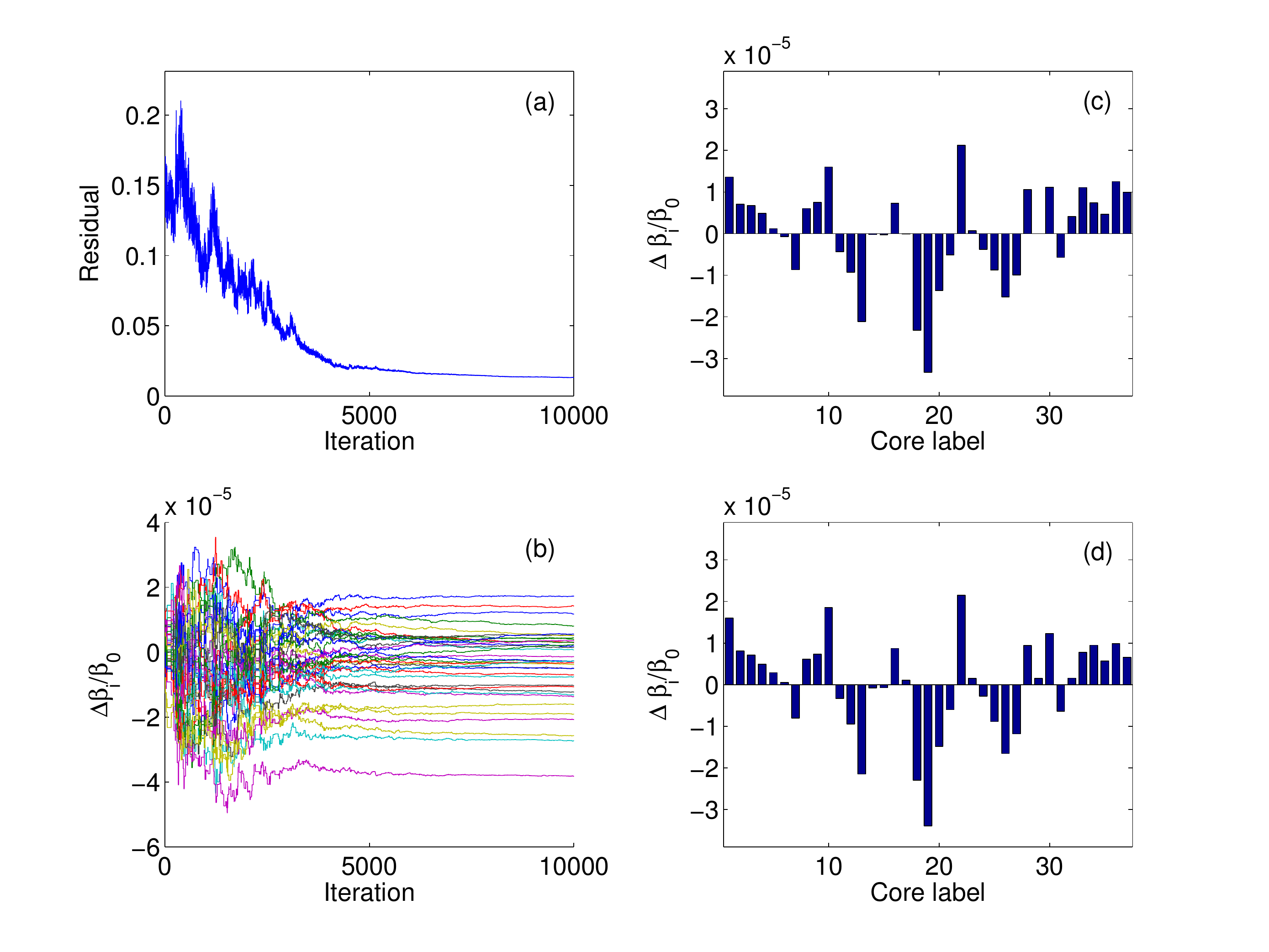}
\caption{Typical performance of simulated reconstruction algorithm on one run. (a) Fit residual $\mathcal{F}$ and (b) intermediate fractional values of $\Delta \beta_i$ as functions of iteration number. (c) Fractional values of target propagation constants $\{\beta_i^\prime\}$ that we aimed to reconstruct. (d) Reconstructed fractional values of propagation constants $\{\beta_i^{\prime \text{(rec)}}\}$.}
\label{fig:sim_rec_1}
\end{figure}

The results for one particular run of the reconstruction are shown in Figure \ref{fig:sim_rec_1}. The reconstruction was performed on a MCF containing 37 cores in a triangular array (a ``3-ring'' MCF). The total run time for the reconstruction was 60 minutes using Matlab on a standard laptop computer. It can be seen that the residual $\mathcal{F}$ gradually decreased as the space of $\{\tilde{\beta}_i^\prime\}$ was sampled, eventually reaching a minimum value. Occasional increases in $\mathcal{F}$ are observed when the algorithm accepts a bad move. The reconstructed values of propagation constant $\{\beta_i^{\prime \text{(rec)}}\}$ are displayed beneath the target set $\{\beta_i^\prime\}$.

\begin{figure}
\centering
\includegraphics[width = 0.8 \textwidth]{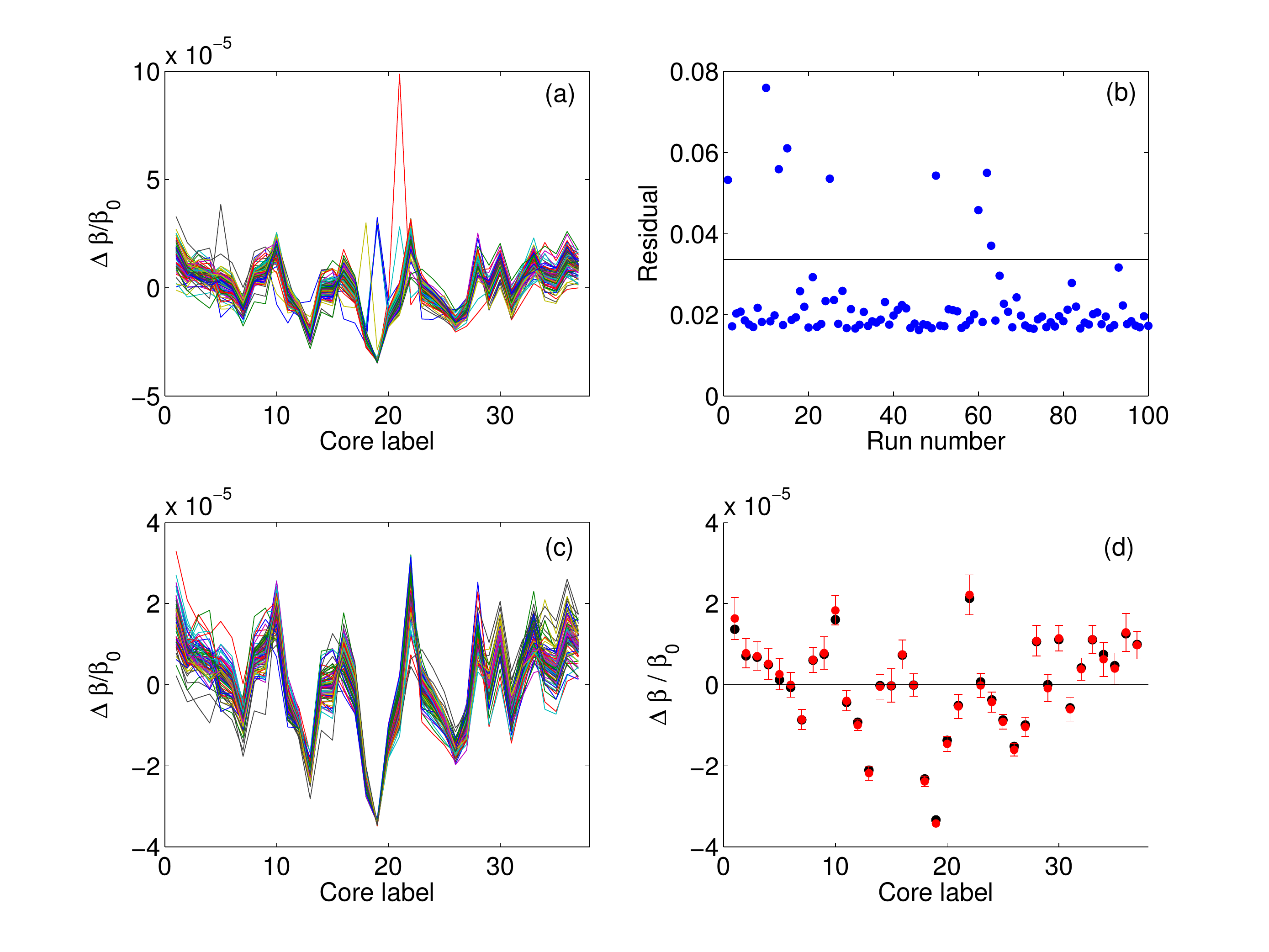}
\caption{Statistics of simulated reconstruction over 100 runs. (a) Raw reconstructed differences in $\beta_i$ for each of 100 runs. (b) Residual value of $\mathcal{F}$, with convergence cut-off illustrated by black line. (c) Reconstructed differences in $\beta_i$ for each of the 91 runs that satisfied convergence condition. (d) Mean reconstructed differences in $\beta_i$ (red) with error bars equal to the standard deviation of the values in (c), with target $\beta_i$ plotted in black.}
\label{fig:sim_rec_100}
\end{figure}

The statistical results over 100 reconstructions of the same set of $\{\beta_i^\prime\}$ are displayed in Figure \ref{fig:sim_rec_100}. This set of reconstructions was carried out in parallel on a quad-core desktop PC, resulting in a  run time of approximately 45 hours. We envisage that this time could be significantly reduced; beyond simple parallelisation of independent reconstructions, no particular effort was expended in optimising the code for speed. 91 of the reconstructions satisfied our convergence condition, defined as having a residual $\mathcal{F}$ within one standard deviation of the mean $\bar{\mathcal{F}}$. The final set of reconstructed propagation constants $\{\bar{\beta}_i^{\prime \text{(rec)}}\}$ plotted in Figure \ref{fig:sim_rec_100} are the mean for each core of the $\{\beta_i^{\prime \text{(rec)}}\}$ values taken from those reconstructions that converged. In the vast majority of cases the target values of $\{\beta_i^\prime\}$ are within the standard deviation of the reconstructed values $\{\bar{\beta}_i^{\prime \text{(rec)}}\}$, confirming the validity of the reconstruction algorithm.

\section{Fabrication and reconstruction of 37-core MCF}

\begin{figure}
\centering
\includegraphics[width = 0.7\textwidth]{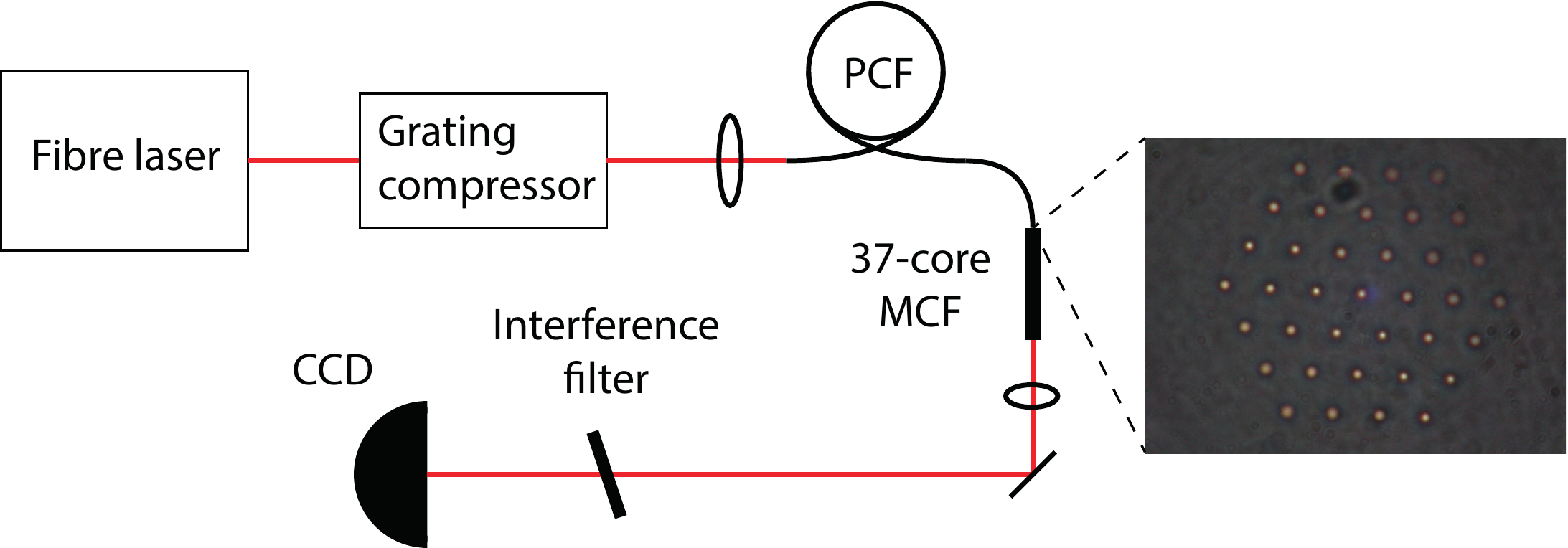} \vspace{5mm}
\includegraphics[width = 0.8\textwidth]{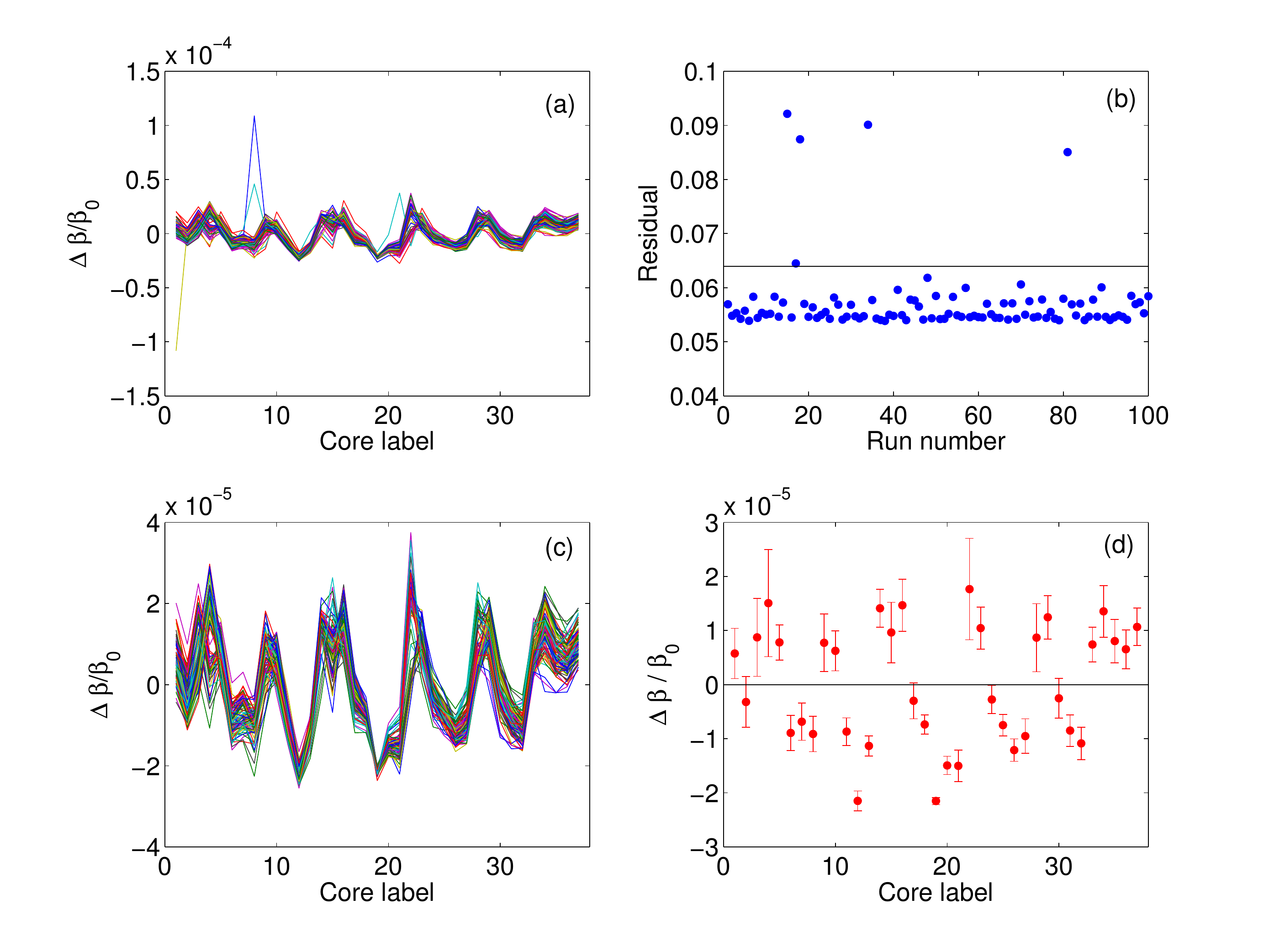}
\caption{(Top) Optical micrograph of the cleaved end face of the 37-core MCF illuminated from below, and schematic of the setup used to characterise the MCF. (Bottom) Results of reconstruction of 37-core MCF over 100 runs. (a) Raw reconstructed differences in $\beta_i$ for each of 100 runs. (b) Residual value of $\mathcal{F}$, with convergence cut-off illustrated by black line. (c) Reconstructed differences in $\beta_i$ for each of the 95 runs that satisfied convergence condition. (d) Mean reconstructed differences in $\beta_i$ (red) with error bars equal to the standard deviation of the values in (c).}
\label{fig:mcf_setup_results}
\end{figure}

We fabricated a 3-ring MCF consisting of 37 cores arranged in a regular triangular array, as shown in Figure \ref{fig:mcf_setup_results}. Each core began as a graded-index Germanium-doped preform that was drawn into a rod, jacketed and re-drawn. The MCF preform was constructed using 37 of these twice-drawn rods stacked with an additional layer of pure silica rods forming a dummy fourth ring to limit the deformation of the third ring of cores during the fibre draw. The MCF preform was drawn to a cane, jacketed, and finally drawn to fibre, resulting in cores with diameters of approximately 1.1\,$\mu$m and separation of 8\,$\mu$m. These cores can be modelled by step-index cores of radius 0.48\,$\mu$m and index contrast 0.02 with an equivalent two-mode beat length of approximately 35\,mm at a wavelength of 650\,nm.

The measurement apparatus is shown in Figure \ref{fig:mcf_setup_results} (a). A 6.9\,mm length of the MCF was cleaved, taking particular care to obtain cleaves that were flat and perpendicular to the fibre axis. Using a femtosecond amplified fibre laser we generated optical supercontinuum in a photonic crystal fibre (PCF) \cite{Stone2008Visibly-white-light} and butt-coupled it directly to the MCF. The output from the MCF passed through a 10\,nm bandpass filter to select the wavelength range of interest and the output face of the MCF was imaged onto a CCD camera. By scanning the PCF across the input face of the MCF we verified that the supercontinuum was coupled into only one core at a time; due to the strong frequency-dependence of the coupling strength, the short-wavelength light reflected from the interference filter allowed us to monitor which core the light was coupled into even when the wavelength reaching the camera had spread across much of the structure. We moved the PCF between all the cores at the input and hence recorded 37 output intensity patterns $\{\pmb{P}_\text{out}^{(k)}\}$ for the 37 possible input conditions $\{\pmb{P}_\text{in}^{(k)}\}$. With the PCF tip at the mid-point between three of the MCF cores, we recorded a background frame to subtract from the data. This was repeated for various wavelength ranges in the visible and near infra-red.

We then ran our reconstruction algorithm on the data measured at a wavelength of 650\,nm. The results of 100 runs are displayed in Figure \ref{fig:mcf_setup_results}. Of these 100 runs, 95 converged and the resulting reconstructed variations in propagation constant for the 37 cores are displayed in panel (d) along with their associated standard deviations. The residual difference $\mathcal{F}$ between the reconstructed and measured intensities for those runs that converged is larger than that for the simulated reconstruction in the previous section for four reasons: noise in the measured data; uncertainty in the uniform value of $g$ used in the reconstruction (though we note that the reconstructed $\{\beta_i^\text{(rec)}\}$ are robust to small errors in $g$); variations in $g$ between cores in the MCF; and the effects of averaging the variation in $g$ over the 10\,nm filter bandwidth. Nevertheless the residual $\mathcal{F}$ for the runs that converged corresponds to a difference between the reconstructed and measured intensities that is approximately a quarter of its value were the propagation constants assumed to be uniform between all the cores, and the output intensities found using $\{\bar{\beta}_i^\text{(rec)}\}$ have less than 6\% of their light in the ``wrong'' cores.

\begin{figure}
\centering
\includegraphics[width = 0.55\textwidth]{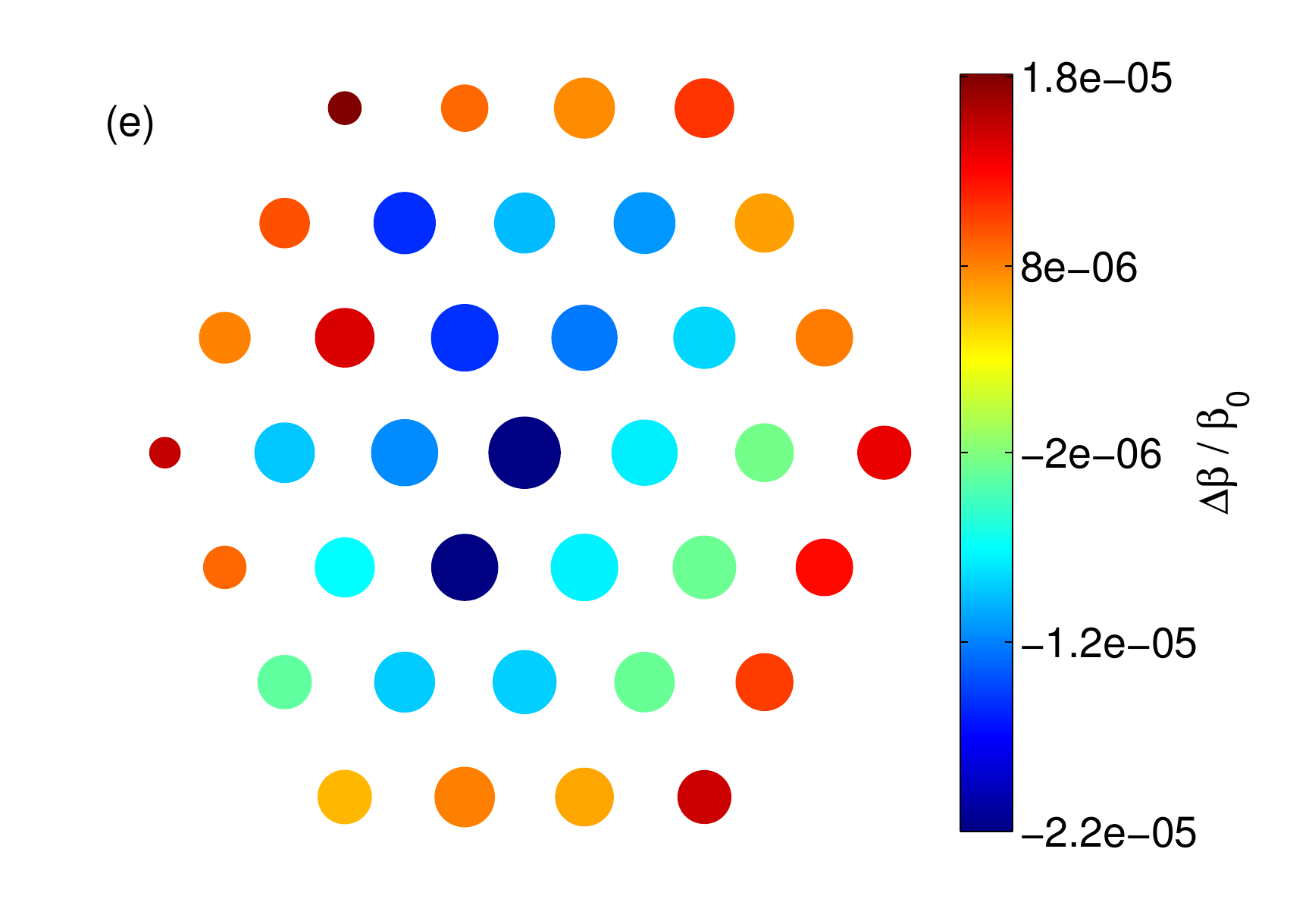}
\caption{Variations in reconstructed $\beta_i$ plotted against position in the MCF structure; size of points represents certainty in reconstructed value.}
\label{fig:structure_results}
\end{figure}

The variations in $\{\bar{\beta}_i^\text{(rec)}\}$ are plotted arranged in their position in the MCF structure in Figure \ref{fig:structure_results}; it can be seen that, although the differences in propagation constant contain a randomly-varying element, there is also a clear pattern to the variations: cores towards the edge of the fibre tend to have a larger propagation constant than those near the centre. This suggests a systematic variation in the structure as the MCF and if this pattern were accounted for only by differences in core size, we see from Figure \ref{fig:two_core} that it would correspond to the outer cores being slightly larger, by a factor of approximately 1.01, than those at the centre. It is also interesting to note that the two cores for which the uncertainty in reconstructed propagation constant is largest are both at the corners of the structure; corner cores have only three nearest neighbours and hence have the smallest interaction with the remainder of the structure.

\section{Conclusion}

We have implemented a robust method of determining the variations in propagation constant in MCF that requires only straightforward measurements of intensity with a simple camera. We outlined an algorithm that allows the variations in propagation constant to be reconstructed if a constant coupling strength is assumed, and demonstrated that the algorithm successfully reconstructs target values of propagation constant. Finally, we applied our technique to a 37-core MCF fabricated in-house and found that the propagation constants vary over a fractional range of $\pm 2\times10^{-5}$.

Our method relies upon the MCF cores being strongly coupled. However, if the cores are similar, there is no fundamental reason why the same technique could not be applied to find the variations in MCF designed to have low coupling strength either by testing it at a longer wavelength or by drawing a section of the fibre preform to a smaller diameter for the purposes of testing. Therefore we anticipate that this method will be of widespread use in characterizing MCF for all application areas.

\section {Acknowledgements}

We gratefully acknowledge support from the UK EPSRC grant EP/K022407/1, the UK STFC grant ST/K00235X/1, and the EU 7$^\text{th}$ Framework Programme under grant agreement 312430.

\end{document}